\documentclass[a4paper]{jpconf}
\usepackage{graphicx}
\usepackage[utf8]{inputenc}
\usepackage[T1]{fontenc}
\usepackage{enumitem}
\usepackage{mathptmx}
\usepackage[svgnames]{xcolor} 
\usepackage{hyperref}         
\usepackage{amsmath}          
\usepackage{amssymb}          
\usepackage{amsfonts}         
\usepackage{amstext}          
\usepackage{textcomp}         
\usepackage{slashed}

\DeclareMathAlphabet{\mathcal}{OMS}{cmsy}{m}{n}

\renewcommand{\bs}[1]{p\!\!\!\!\! p \!\!\!\!\! p\!\!\!\!\! p}

\begin{document}
\title{Towards the dynamical study of heavy-flavor quarks in the Quark-Gluon-Plasma}

\author{H.~Berrehrah$^1$, E.~Bratkovskaya$^1$, W.~Cassing$^2$, P.B.~Gossiaux$^3$ and J.~Aichelin$^3$}

\address{$^1$ Frankfurt Institute for Advanced Studies and Institute for Theoretical Physics, Johann Wolfgang Goethe Universit\"at, Ruth-Moufang-Strasse 1, 60438 Frankfurt am Main, Germany\\
$^2$ Institut f\"ur Theoretische Physik, Universit\"at Giessen, 35392 Giessen, Germany\\
$^3$ Subatech, UMR 6457, IN2P3/CNRS, Universit\'e de Nantes, \'Ecole des Mines de Nantes, 4 rue Alfred Kastler, 44307 Nantes cedex 3, France}

\ead{berrehrah@fias.uni-frankfurt.de}

\begin{abstract}
  Within the aim of a dynamical study of on- and off-shell heavy quarks $Q$ in the quark gluon plasma (QGP) - as produced in relativistic nucleus-nucleus collisions -  we study the heavy quark collisional scattering on partons of the QGP. The elastic cross sections $\sigma_{q,g-Q}$ are evaluated for perturbative partons (massless on-shell particles) and for dynamical quasi-particles (massive off-shell particles as described by the dynamical quasi-particles model ``DQPM'') using the leading order Born diagrams. We demonstrate that the finite width of the quasi-particles in the DQPM has little influence on the cross sections $\sigma_{q,g-Q}$ except close to thresholds. We, furthermore, calculate the heavy quark relaxation time as a function of temperature $T$ within the different approaches using these cross sections.
\end{abstract}

\section{Introduction} 

   The heavy quarks $Q$ are created in hard processes in relativistic heavy-ion collisions (HIC) and are initially isotropically distributed in the transverse momentum space. To explain the final azimuthal anisotropy observed in experiments, one has to study in detail the dynamics and the transport properties of the heavy quarks in the medium, i.e. in the quark gluon plasma (QGP). Since the experimental discovery of the strongly interacting QGP (sQGP), many models and theoretical schemes have been set up to describe the large coupling of heavy quarks to the hot QGP. Indeed, the early studies of $Q$ scattering in the QGP based on pQCD cross sections failed to achieve a quantitative understanding of the experimental data \cite{Gossiaux:2008jv}. The purpose of this contribution is to explore new descriptions for the interaction between heavy quarks and the QGP partons ($q,g$) at   finite temperature. Therefore, we will present the corresponding elastic cross sections following different models and we will end by evaluating the heavy quark relaxation time as a function of the temperature $T$ of the medium.

\vspace*{-0.2cm}
\section{Cross sections at finite temperature} 

   The central question about the heavy quark (Q) dynamics is the magnitude and detailed form of the interactions between the quark $Q$ and the partons ($q,g$) of the finite temperature medium. Indeed, the evaluation of the corresponding cross sections depends on the (running) coupling  in the interaction vertices and the quark and gluon propagators at finite $T$. Although the scattering of heavy quarks in vacuum and in the QGP for massless partons has extensively been studied in the literature \cite{Combridge1979429}   using pQCD, a full treatment of finite temperature effects is still needed to consider the case of  non-perturbatively interacting massive light and heavy quarks and gluons.  In this contribution, we present an effective approach for the derivation of the on- and off-shell cross sections for the interaction of massive dynamical quasi-particles as constituents of the finite temperature strongly interacting medium (sQGP). In order to consider all the effects of non-perturbative nature of the sQGP constituents, we refrain from a fixed-order thermal loop calculation  and pursue instead a more phenomenological approach. The multiple strong interactions of quarks and gluons in the sQGP here are encoded in their effective propagators-extracted from lattice QCD data within the DQPM \cite{Wcassing2009EPJS}- with dynamical spectral functions and an effective running coupling $g^2(T)$ which is enhanced in the infrared.

%

\vspace*{-0.2cm}
\section{$qQ$ and $gQ$ elastic scattering}

The matrix elements for the elastic scattering of a heavy quark with a light quark $qQ \rightarrow qQ$ (resp. gluon $gQ \rightarrow gQ$) have been calculated for the case of massless partons in Ref. \cite{Combridge1979429}. These pQCD cross sections have to be supplemented by two parameters to allow for a quantitative evaluation: the coupling constant $\alpha_s$ and the infrared (IR) regulator $\mu$ which renders the cross section infrared finite. The value of the cut-off or of the gluon mass has been fixed according to phenomenological considerations. Therefore, the uncertainties in these scales is large which also leads to corresponding uncertainties in the cross sections. In our study we will present explicit values for these two parameters which are based on different theoretical considerations. Moreover, we will take into account the quasi-particle nature of the incoming and outgoing particles by incorporating spectral functions. In this way we will be able to test to what extent the quasi-particle nature of quarks and gluons will influence the heavy quark scattering. We briefly recall  the ingredients of the different approaches studied in this paper:

\begin{itemize}[leftmargin=* ,parsep=0cm,itemsep=0cm,topsep=0cm]
\item \textbf{HTL-GA}: from Gossiaux and Aichelin following a Hard-Thermal-Loop inspired approach \cite{Gossiaux:2008jv,Berrehrah:2013mua} for massless light quarks and gluons. As compared to former pQCD approaches this model differs in the description of $q,g\!-\!Q$ interactions in two respects:

  \begin{itemize}[leftmargin=* ,parsep=0cm,itemsep=0cm,topsep=0cm]
  \item an effective running non-perturbative coupling,  which remains finite for vanishing 4-momentum transfer squared  $t \rightarrow 0$, is employed \cite{Dokshitzer:1995qm},
  \item an infrared regulator in the $t$-channel, which is determined by hard thermal loop calculations. The $t$-channel is adjusted to give the same energy loss as calculated in a HTL approach \cite{Gossiaux:2008jv}.
  \end{itemize}

 In the HTL-GA approach we will also consider the case with fixed coupling $\alpha_s$ and the Debye mass ($m_D \approx \xi g_S T$) as an infrared regulator.

\item \textbf{IEHTL} (Infrared Enhanced Hard Thermal Loop): the main characteristics of this approach are:

\begin{itemize}[leftmargin=* ,parsep=0cm,itemsep=0cm,topsep=0cm]
\item off-shell kinematics, in particular light quarks and gluons virtualities,
\item internal and external lines of light and heavy quarks as well as gluons are dressed with non-perturbative spectral functions,
\item parametrizations of the quark and gluon propagators from the DQPM that are matched to reproduce lQCD data in equilibrium,
\item $qgq$ and $QgQ$ vertices are modified compared to pQCD vertices by replacing the perturbative coupling with the full running coupling depending on temperature according to the DQPM.
\end{itemize}

\item \textbf{DpQCD} (Dressed pQCD): in the same spirit as the IEHTL model with the difference that for DpQCD we only consider the DQPM pole masses for the QGP partons thus treating the partons on-shell.
\end{itemize}

\begin{figure}[h!] 
\begin{minipage}{17pc}
\begin{center}
\includegraphics[width=16pc, height=14.5pc]{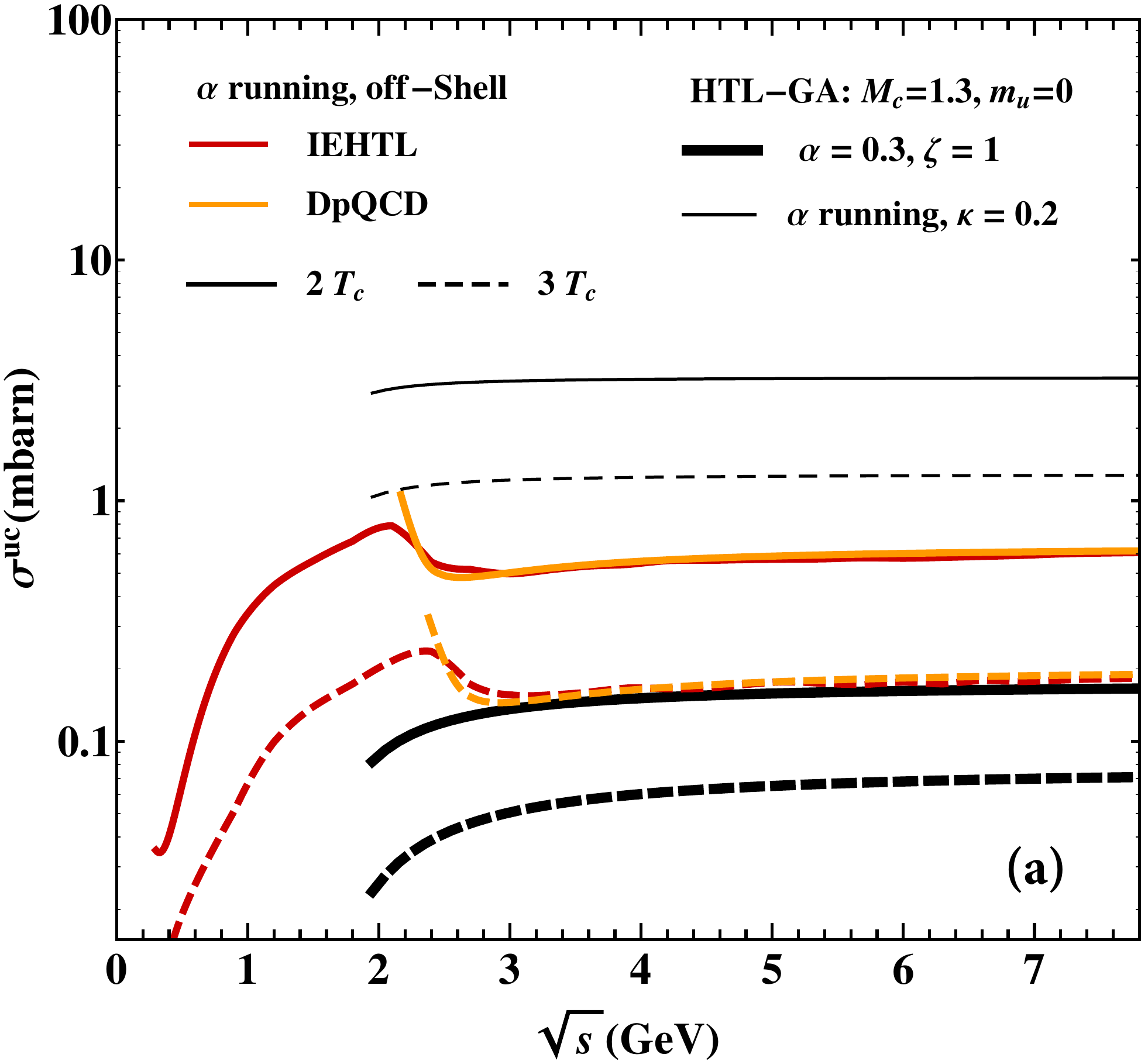}
\end{center}
\end{minipage}\hspace{1.5pc}
\begin{minipage}{17pc}
\begin{center}
\includegraphics[width=16pc, height=14.45pc]{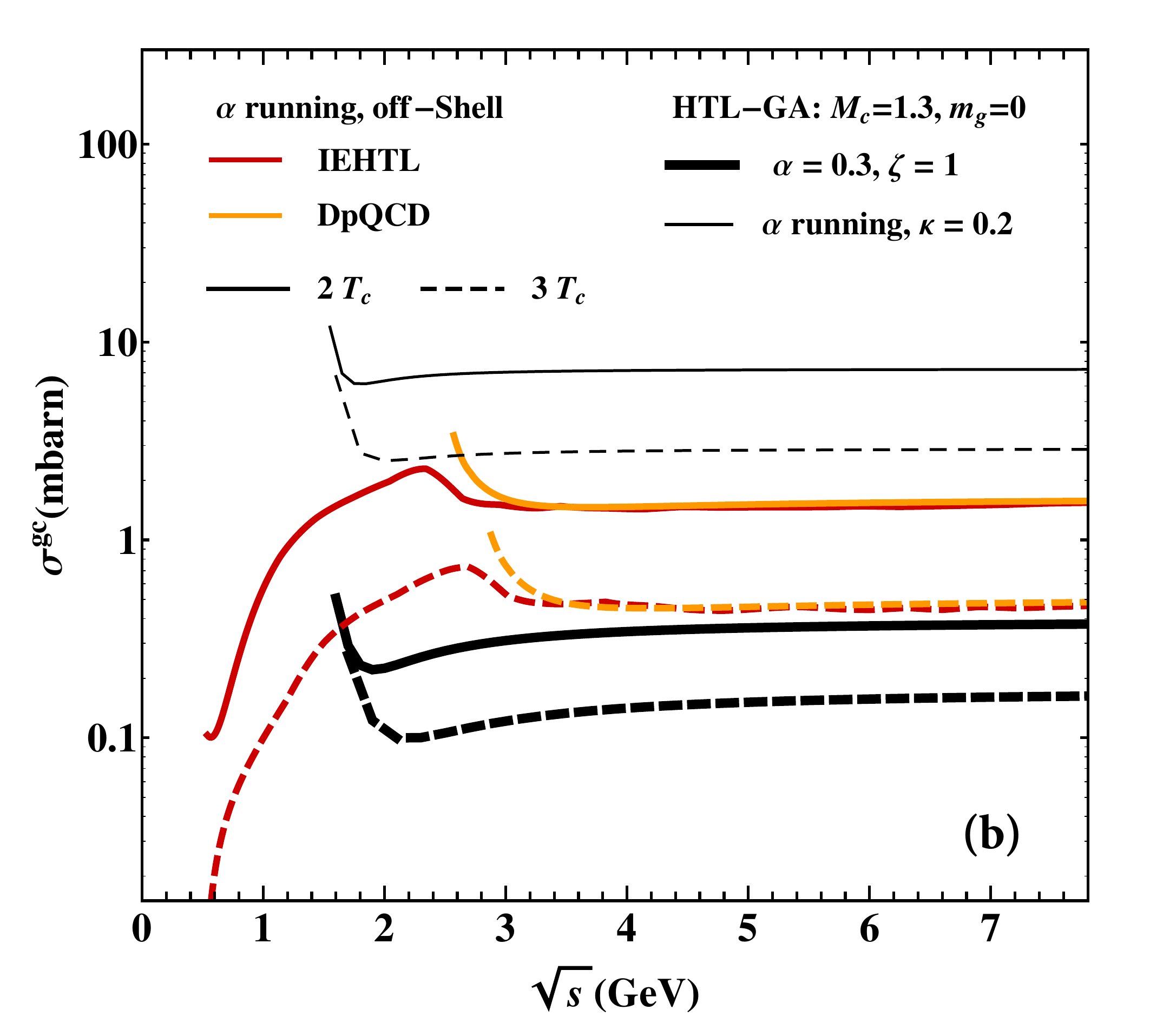}
\end{center}
\end{minipage}
\vspace{-0.2cm}
\caption{\emph{Comparison of $\sigma_{u,g-c}$ calculated within the HTL-GA (black lines), DpQCD (yellow lines)
and IEHTL (red lines) approaches as a function of $\sqrt{s}$ for different temperatures $T$ (see legend).
$m_{u,g}$ and $M_c$ are given in GeV. Left~ (a): $u c \rightarrow u c$ ; Right~ (b): $g c \rightarrow g c$.}}
\label{fig:SigmagQDpQCD-HTLapproach}
\end{figure}

  Fig. \ref{fig:SigmagQDpQCD-HTLapproach} (a) (resp. (b)) present the $qQ$ (resp. $gQ$) elastic scattering cross sections calculated within the models presented above as a function of $\sqrt{s}$ for different temperatures ($2 T_c, 3 T_c$). Apart from threshold effects $\sigma_{q,g-Q}$ are approximately independent on $\sqrt{s}$, however, differ substantially in magnitude. These figures show clearly the different values for $\sigma_{q,g-Q}$ for the different choices of couplings and infrared regulators $\mu$. Since the values of $\mu$ increase with increasing temperature, all the cross sections decrease for the temperature of $3 T_c$ (dashed lines) in comparison to the $2 T_c$ case (solid lines).

 The comparison of the DpQCD/IEHTL approaches demonstrates that the off-shell mass distributions only have a sizeable impact close to the threshold given by the pole masses. This is due to the moderate parton widths considered in the DQPM model. At energies below the on-shell threshold the off-shell cross section increases with $\sqrt{s}$ because more and more masses can contribute. Whereas the on-shell cross section diverges at the threshold the off-shell cross section shows a maximum at the on-shell threshold and decreases then due to the decrease of the on-shell cross section.

\vspace*{-0.8cm}
\section{Transition rate and relaxation time} 

To compute transport coefficients one needs first to evaluate the elastic transition rate $\omega (T)$ which is given for the case of on-shell ($\omega_{jQ}^{\textrm{on}}$) and off-shell ($\omega_{jQ}^{\textrm{off}}$) partons, with $j = q$, by \cite{Marty:2013ita}:

\vspace{-0.6cm}
{\setlength\arraycolsep{0pt}
\begin{eqnarray}
& &   \omega_{qQ}^{\textrm{on}} (m_q^i, M_Q^i,m_q^f, M_Q^f, T) = \int \limits_{\text{Th}}^\infty
d s \ \sigma_{qQ}^{\textrm{on}} (m_q^i, M_Q^i,m_q^f, M_Q^f,T,s) \ \mathcal{P}^{\textrm{on}}(m_q^i, M_Q^i,T,s),
\nonumber
\end{eqnarray}}
\vspace{-1.1cm}
{\setlength\arraycolsep{0pt}
\begin{eqnarray}
\hspace*{-5.8cm} \omega_{qQ}^{\textrm{off}} (T) = \int \limits_{\text{Th}}^\infty d s \
\sigma_{qQ}^{\textrm{off}} (T,s) \ \mathcal{P}^{\textrm{off}}(T,s),
\label{equ:1}
\end{eqnarray}}
\vspace{-0.5cm}

 with the threshold $\text{Th} =$ max\{$(m_q^i + M_Q^i)^2$,$(m_q^f + M_Q^f)^2$\} and $\mathcal{P}$ denoting the probability for a $qQ$ pair with the energy $\sqrt{s}$ in the medium at finite $T$:

\vspace{-0.55cm}
{\setlength\arraycolsep{0pt}
\begin{eqnarray}
& & \mathcal{P}^{\textrm{on}}(m_q^i, M_Q^i,T,s) = C^{\textrm{on}} \frac{E_1 E_2}{\sqrt{s}} p_{cm} (s) v_{rel} (s) f(E_1)f(E_2)
\nonumber\\
& & {}  \mathcal{P}^{\textrm{off}}(T,s) = \int d m_q^i \int\limits d M_Q^i \
\frac{E_1 E_2}{\sqrt{s}} \ C^{\textrm{off}}(T) \ p_{\text{cm}}(s) v_{rel} (s) \ f_q\left(E_1 \right)
f_{Q}\left(E_2 \right) A_{q^i} (m_q^{i}) A_{Q^i} (M_Q^{i}),
\label{equ:2}
\end{eqnarray}}
\vspace{-0.5cm}

  with the center-of-mass momentum $p_{\text{cm}}$, relative velocity $v_{rel}$ and the normalization factors $C^{\textrm{on}}$ and $C^{\textrm{off}}$. In (\ref{equ:2}) $f_{q,Q} (E)$ is the Fermi-Dirac distribution for the light and heavy quark. Similar calculations can be performed for massive gluons by including the Bose-Einstein
distribution in Eq. (\ref{equ:2}).

\begin{figure}[h!]
\hspace*{-0.5cm}\includegraphics[width=13.3pc, height=14.7pc]{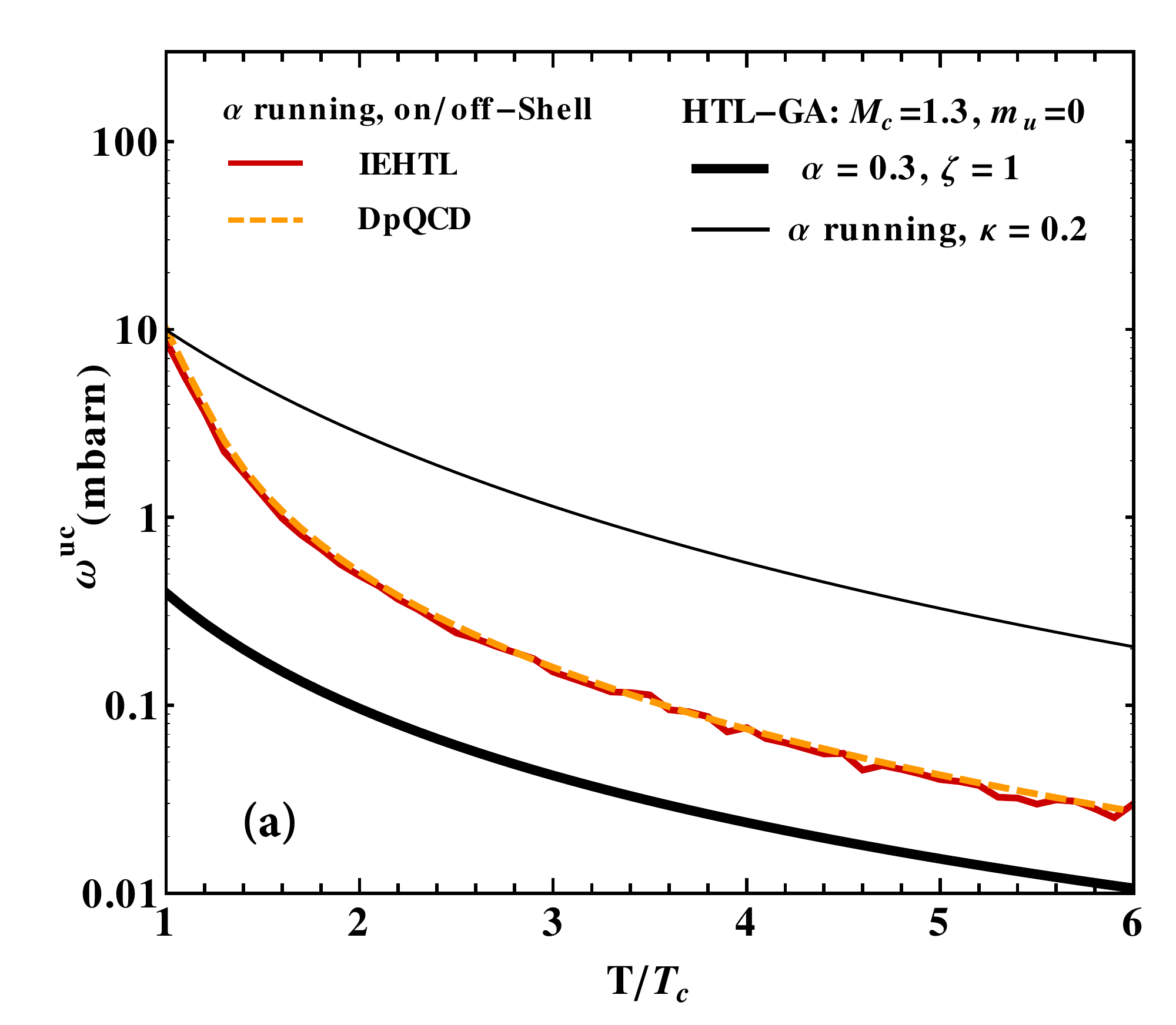}
\hspace*{-0.09cm}\includegraphics[width=13.3pc, height=14.7pc]{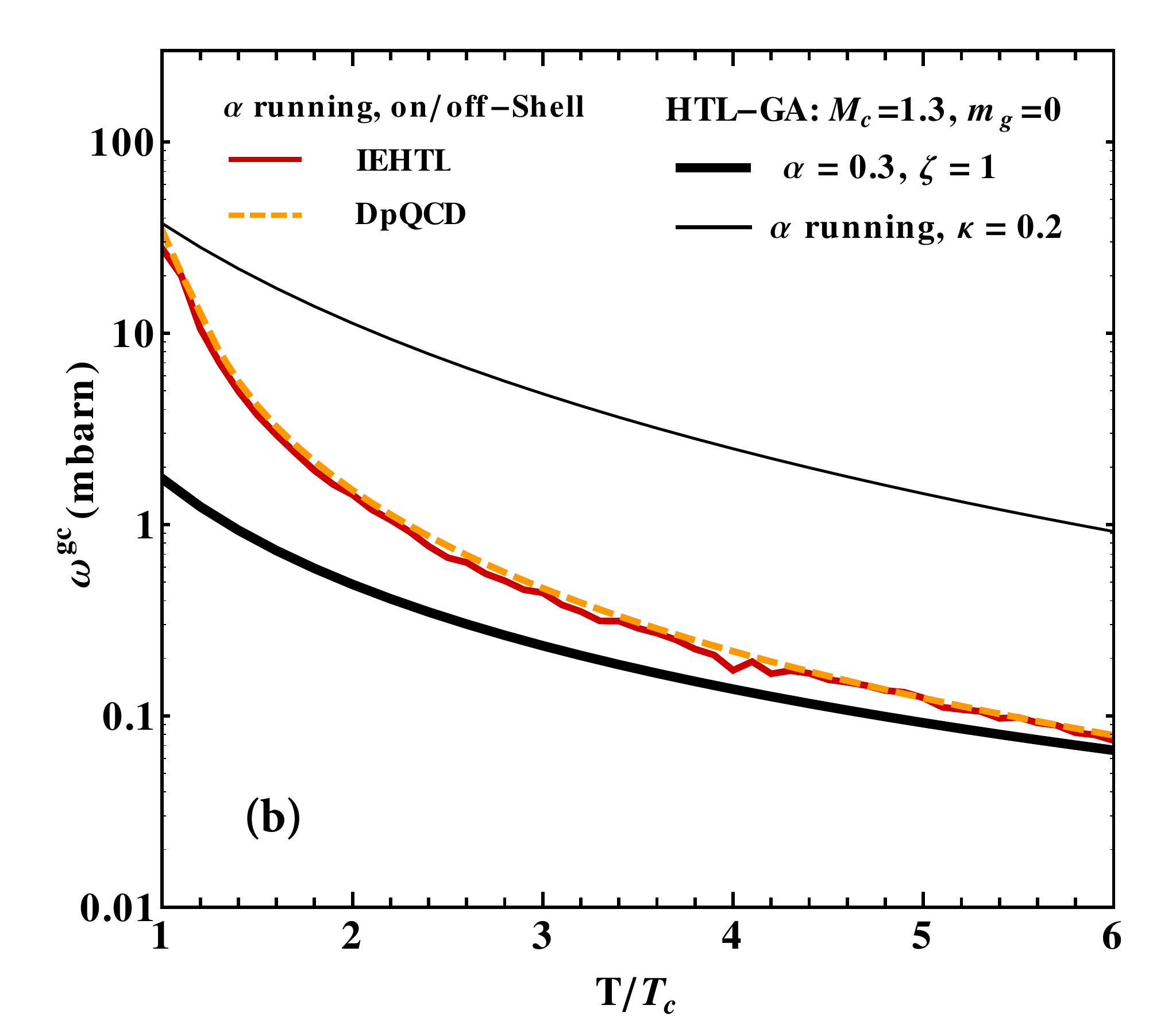}
\hspace*{-0.09cm}\includegraphics[width=13.4pc, height=14.7pc]{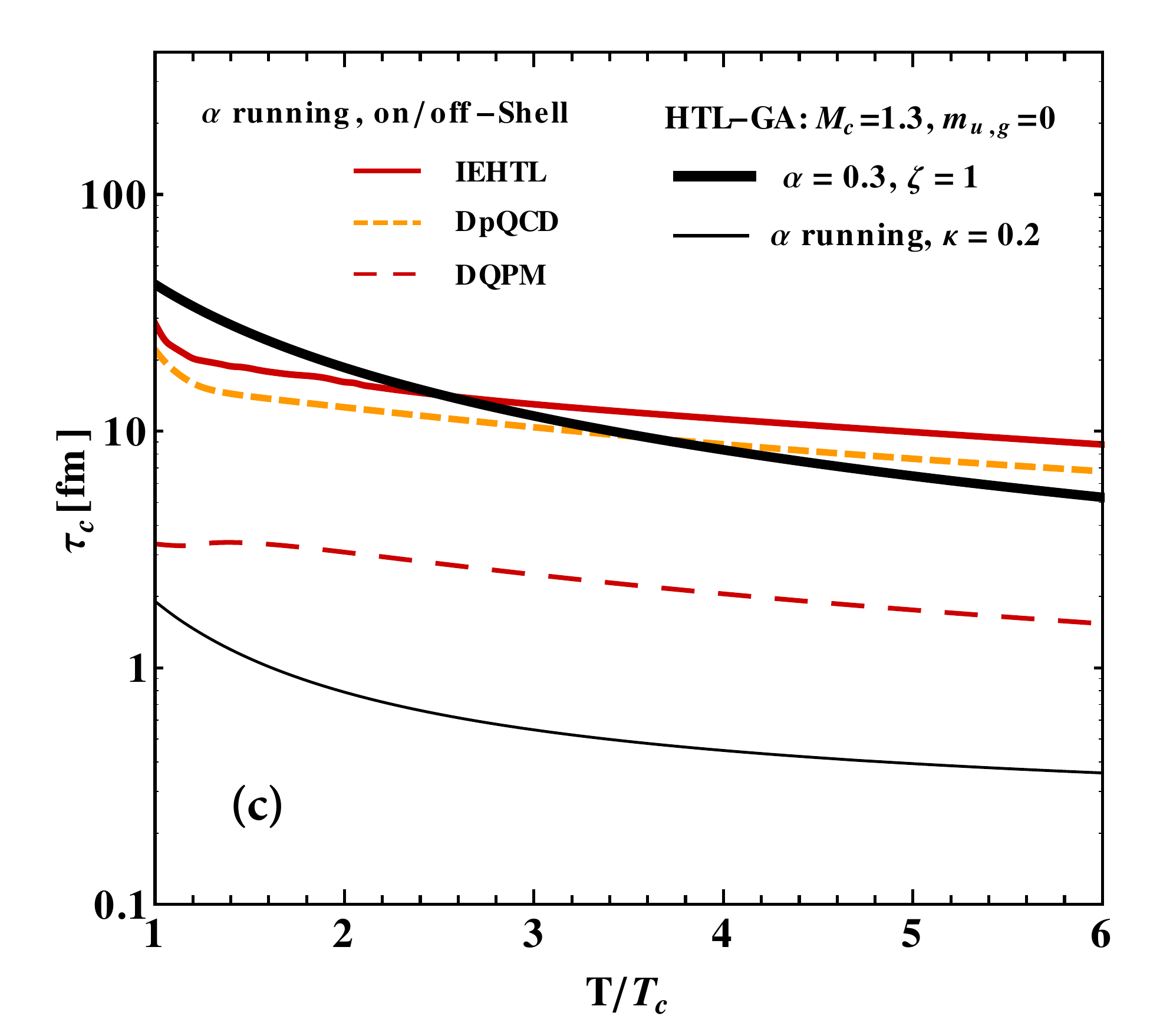}
\vspace{-0.7cm}
\caption{\emph{$uc$ (a) and $gc$ (b) elastic transition rate, (c) heavy quark relaxation time as a function of $T/T_c$ with $T_c = 158$ MeV. Presented here are different models for off-shell scattering with the DQPM spectral functions (IEHTL (DQPM), red silde (dashed) line) and on-shell partons with the DQPM pole masses (DpQCD, yellow dashed lines). Also shown is a comparison to results of the HTL-GA approaches with constant coupling (thick black line) and running coupling (thin black line), where $M_c$ is given in GeV.} \label{fig:IntegratedSigmaHTL-DpQCD-IEHTL}}
\vspace*{-0.5cm}
\end{figure}

  Figure \ref{fig:IntegratedSigmaHTL-DpQCD-IEHTL}-(a) (resp. (b)) shows the $qQ$ (resp. $gQ$) elastic transition rate as a function of the medium temperature $T/T_c$ for the different approaches presented above. We notice different power laws in $T$, i.e. $\sim \displaystyle T^{-\beta}$ for the HTL-GA and DpQCD/IEHTL models. In fact, one can find that ($\displaystyle \beta^{T<1.2 T_c} \sim 2, \beta^{T>1.2 T_c} \sim 1.7$) for the HTL-GA versions, whereas ($\displaystyle \beta^{T<1.2 T_c} \sim 4, \beta^{T>1.2 T_c} \sim 2$) for the DpQCD/IEHTL approaches that are practically identical. The higher power coefficients in the DpQCD/IEHTL approaches can be traced back to the infrared enhancement of the effective coupling. These different power laws in $T$ will have a sizeable effect on the transport coefficients. We stress again that the effect of the DQPM spectral function on $\displaystyle \omega (T)$ is negligible by comparing $\omega (T)$ using the DpQCD/IEHTL approaches.

 By means of the transition rates we can compute the relaxation time $\tau$ for particles within the different models. In the dilute gas approximation the relaxation time $\tau$ is obtained for on-shell particles $(\tau_c^{-1})_{DpQCD}$,$(\tau_c^{-1})_{HTL-GA}$ and for off-shell quasi-particles $(\tau_c^{-1})_{IEHTL}$,$(\tau_c^{-1})_{DQPM}$ by (\ref{equ:3}) \cite{Marty:2013ita}, where $n_{i}$ is the quark/antiquark or gluon density.  For the DQPM we do not need the explicit cross sections since the inherent quasi-particle width $\gamma_c (T)$ directly provides the total interaction rate \cite{Wcassing2009EPJS}.

\vspace{-0.7cm}
{\setlength\arraycolsep{-1pt}
\begin{eqnarray}
\label{equ:3}
& & \displaystyle(\tau_c^{-1})_{DpQCD} = \!\!\!\!\! \sum\limits_{i \in{q,\bar{q},g}} \!\!\!\! n_i^{\textrm{on}} (T) \
\omega_{\textrm{i c}}^{DpQCD} (T), \hspace{0.4cm} \displaystyle(\tau_c^{-1})_{HTL-GA} = \!\!\!\!\!
\sum\limits_{i \in{q,\bar{q},g}} \!\!\!\! n_i^{\textrm{on}} (T) \ \omega_{\textrm{i c}}^{HTL-GA} (T),
\hspace{0.3cm} \displaystyle (\tau_c^{-1})_{DQPM} = \frac{\hbar c}{\gamma_c (T)}
\nonumber\\
& & {} \displaystyle(\tau_c^{-1})_{IEHTL} = \!\! \sum\limits_{i \in{q,\bar{q},g}} n_i^{\textrm{off}} (T)
\ \omega_{\textrm{i c}}^{IEHTL} (T), \hspace{0.5cm} \displaystyle n_i^{\textrm{off}} (T) = \int \!\!\!\! \int
\!\! \frac{d^3 p}{(2 \pi)^3} A_i^{BW} (m_i) d m_i \ f_i\!/\!g_i (p, T, m_i,\mu_i).
\end{eqnarray}}
\vspace{-0.24cm}

  Fig. \ref{fig:IntegratedSigmaHTL-DpQCD-IEHTL}-(c) shows the heavy quark relaxation time for the case of on-shell and off-shell partons at finite $T$ following the DpQCD/IEHTL and HTL-GA approaches. We deduce that the heavy quark needs more interactions to relax when its interactions are described by $\sigma^{HTL-GA}$ as compared to  $\sigma^{DpQCD/IEHTL}$.

\vspace*{-0.2cm}
\section{Summary}

  We have presented the elastic scattering of heavy quarks with quarks and gluons in a quark-gluon-plasma using two approaches based on different regularization schemes. $i)$ The dynamical quasi-particle model (DQPM) \cite{Wcassing2009EPJS,Berrehrah:2013mua} in which quarks and gluons have a finite mass and width that vary with temperature $T$. $ii)$ The Peshier-Gossiaux-Aichelin approach \cite{Gossiaux:2008jv} which uses massless partons, a running coupling $\alpha_s(Q^2)$ and an infrared regulator which have been adjusted to reproduce the heavy quark energy loss in heavy-ion collisions at RHIC energies. The size of the elastic cross sections is dominated by the infrared regulator which in the finite temperature medium is determined by a dynamical gluon mass.

  We have demonstrated that the finite width of the partons in the DQPM has little influence on the cross sections except close to thresholds. The total cross sections show a smooth dependence on the invariant energy. Even if the heavy quark relaxation time shows large differences between the different models, explicit transport calculations in comparison to experimental data are needed to figure out the appropriate scenario.

\vspace*{-0.2cm}
\section*{References}
\bibliographystyle{iopart-num}
\bibliography{References}

\end{document}